\documentclass{article}
\usepackage{amsmath,amsthm}
\theoremstyle{plain}
\newtheorem{thm}{Theorem}

\theoremstyle{remark}

\theoremstyle{definition}

\begin{document}

\title{On Lie Point Symmetries of Einstein's equations for the Friedmann-Roberstson-Walker Cosmology.}
\author{Paschalis G. Paschali\footnote{email:pashalis@cytanet.com.cy}\\
Department of Mathematics and Statistics\\
University of Cyprus\\
Nicosia, P.O.Box 537\\
Cyprus\\
Georgios C. Chrysostomou\footnote{email: eng.cg@fit.ac.cy}\\
Frederick University, Pallouriotissa,
1036 Nicosia, Cyprus}
\date{\today}

\maketitle

\begin{abstract}
We study the Lie point symmetries of Einstein's equations for the Friedmann-Roberstson-Walker Cosmology. 
They form either a two - dimensional or a three - dimensional solvable group depending 
on the form of the self interacting potential. Using the invariants of the 
group we reduce the second order system of differential equations into a first order system. 
Writing the action in terms of the proper time we study the point symmetries and the variational symmetries 
of the resulting equations. 
\end{abstract} 

\subsubsection*{Keywords}
Lie Point Symmetries, Friedmann-Roberstson-Walker Cosmology, solvable group, Noether's theorem

\section{Introduction}
\label{s:intro}
 A symmetry group of a system of differential equations is a Lie group acting on the space of independent 
 and dependent variables in such a way, that solutions are mapped into other solutions. Knowing the symmetry 
 group allows one to determine some special types of solutions that  are invariant under a subgroup of the full 
 symmetry group, and in some cases one can solve the equations completely. The symmetry approach to solving 
 differential equations can be found, for example, in the books of Olver [10], Bluman and Cole [1], Bluman and 
 Kumei [2], Fushchich and Nikitin [5] and Ovsiannikov [11]. 
 
Einstein's General Theory of Relativity is based in the most fundamental way on the concepts of symmetry and 
general covariance. Lie's Theory on the other hand is the most systematic mathematical way to study symmetries 
of differential equations. So it is obviously interesting to apply the Lie methods to a theory fundamentally 
related with the idea of symmetry and general covariance. 
In this paper we study the Lie point symmetries of the notorious Einstein's equations 
for the Friedmann-Roberstson-Walker Cosmological model with a matter field of self interacting potential $V(\phi)$. 
For completeness we include here their derivation from the Einstein-Hilbert action. For the above Cosmological model it is given by 

\begin{equation}
S=\frac {1}{2}\int dt \frac {N}{a} \Big[-\Big(\frac {a}{N} \frac {da}{dt}\Big)^2 +ka^2 + 
\Big(\frac {a^2}{N} \frac {d \phi}{dt}\Big)^2 - 2a^4 V\Big]
\end{equation} 
where N is the lapse function and $a$ is the scale factor or the size of the Universe, and $\phi$ is the scalar 
field of self interacting potential $V(\phi)$. If we vary the action $S$ with respect to $N$, 
$\phi$ and $a$ we will get the following equations:  
\begin{equation}
\dot{a}^2 = 2a^2 V +a^2 \dot {\phi}^2 -k
\end{equation}
\begin{equation}
0 = \ddot{\phi} +3\frac{\dot{a}}{a}\dot{\phi} +\frac{dV}{d \phi}
\end{equation}
\begin{equation}
0=-2a\ddot{a}-\dot{a}^2-k-3a^2\dot{\phi}^2 +6a^2V
\end{equation}
where the dot above means $\frac{1}{N}\frac{d}{dt}$. Since we have imposed the Hamiltonian constraint 
we can set $N=1$. Using Equations (2) and (3), Equation (4) becomes: 
\begin{equation}
\ddot{a} = 2a V -2a \dot {\phi}^2
\end{equation} 
which is usually called the second Einstein equation. Here $k=+1$ for the case of a closed universe for 
which the space part is a three sphere, $k=0$ for the case of a marginally open universe with 
flat space part and finally $k=-1$ represents an open universe whose space part is a three dimensional 
hyperboloid embedded in Minkowski space. 

The above equations are not only important for classical cosmology but they are also important for the 
semiclassical approximation to the Wheeler-DeWitt equation [4], i.e., to the semiclassical quantum cosmology.  
The Wheeler-DeWitt equation for the Friedmann-Roberstson-Walker Cosmology takes the form: 
\begin{equation}
\Big(-\frac{1}{2}\nabla^2 +U \Big)\Psi =0
\end{equation}
where the Laplacian operator is calculated in the minisuperspace metric 
\begin{equation}
ds^2 =-ada^2+a^3d\phi^2
\end{equation}
and the potential is given by 
\begin{equation}
U=-\frac{a}{2}+a^3V
\end{equation}
If we write the wavefunction of the universe in the form 
\begin{equation}
\Psi \simeq Ce^{iS}
\end{equation}
where $C$ is slowly varying with respect to $S$, then in the W.K.B. approximation we choose 
the action to satisfy the Hamilton-Jacobi equation 
 \begin{equation}
\frac {1}{2} (\nabla S)^2-k \frac {a}{2}+a^3 V=0
\end{equation}
which is consistent with the fact that $\Psi$ satisfies the Wheeler-DeWitt equation. The action $S$ 
defines then a vector field 
\begin{equation}
\frac {d}{dt}=\vec {\nabla}S \cdot \vec {\nabla}
\end{equation}
It can be easily proved that its trajectories satisfy the above classical equations (2), (3) and (5). 
See for example [12]. In a forthcoming paper we study the symmetries of the Wheeler-DeWitt and of the 
Hamilton-Jacobi equations and we compare them with the results of this work. 
Finally we note that the quantity 
$\dot {a}^2 - a^2 \dot {\phi}^2 -2a^2 V(\phi)$ 
is a constant of motion for the second order equations (3) and (5), which we study next. 

\section{Study of Lie point symmetries}

We study here the Lie point symmetries of the equations 
\begin{equation}
\ddot {a} = 2aV(\phi)-2a\dot {\phi}^2
\end{equation}
 \begin{equation}
\ddot {\phi} =-3 \frac{\dot {a}}{a} \dot {\phi} - \frac{dV}{d\phi}
\end{equation}
where $V(\phi)$ is an arbitrary function of $\phi$. 
Here we give some details of the calculations, because later we apply the same method to equations (3) 
and (4) without showing any calculations. 
If the vector field $G$ is given by 
 \begin{equation}
G=\tau \frac {\partial}{\partial t}  +A \frac {\partial}{\partial a}  + \Phi \frac {\partial}{\partial \phi}
\end{equation} 
then its second prolongation or extension is given by 
\begin{equation} 
\begin{split} 
pr^{(2)} G=&\tau \frac {\partial}{\partial t}  +A \frac {\partial}{\partial a}  + \Phi \frac {\partial}{\partial \phi} + (\dot {A} - \dot {\tau}\dot{a})\frac {\partial}{\partial \dot{a}} + (\dot {\Phi} - \dot {\tau}\dot{\phi})\frac {\partial}{\partial \dot{\phi}} \\ 
&+ (\ddot {A} - \ddot {\tau}\dot{a}- 2\dot{\tau} \ddot {a})\frac {\partial}{\partial \ddot{a}}
+ (\ddot {\Phi} - \ddot {\tau}\dot{\phi}- 2\dot{\tau} \ddot {\phi})\frac {\partial}{\partial \ddot{\phi}}
\end{split} 
\end{equation}
In the above equations $\tau$, $A$ and $\Phi$ are functions of $t$, $a$ and $\phi$. 
The necessary and sufficient condition for $G$ to be an infinitesimal generator of the 
symmetry group of the above equations (12) and (13) are: 
\begin{equation} 
pr^{(2)} G(Eq12)=0
\end{equation} 
\begin{equation} 
pr^{(2)} G(Eq13)=0
\end{equation} 
or equivalently 
\begin{equation} 
\ddot{A}-\ddot{\tau}\dot{a}-2\dot{\tau}\ddot{a}=2AV+2A\Phi \frac{dV}{d\phi}-2A\dot{\phi}^2 
-4a\dot{\phi}(\dot{\Phi}-\dot{\tau}\dot{\phi})
\end{equation} 
\begin{equation}
\begin{split}
\ddot {\Phi} - \ddot {\tau} \dot{\phi} -2 \dot {\tau} \ddot{\phi} & = -3(\dot{A}-\dot{\tau}\dot{a}) \frac{\dot{\phi}}{a}
-3(\dot{\Phi}-\dot{\tau}\dot{\phi}) \frac{\dot{a}}{a}\\
& +3\frac{\dot{a}\dot{\phi}}{a^2}A-\frac{d^2 V}{d\phi^2}\Phi\\
\end{split}
\end{equation}
Expanding Equation (18) we get 
\begin{equation}
\begin{split}
&A_{tt} +2A_{ta}\dot{a}+2A_{t\phi} \dot{\phi} +2A_{a\phi }\dot{a}\dot{\phi}+A_{aa}\dot{a}^2 +A_{\phi\phi} \dot{\phi}^2 +2aV(\phi)A_a \\
&- 2a\dot{\phi}^2 A_a -3\frac{\dot{a}\dot{\phi}}{a}A_\phi -\frac{dv}{d\phi}A_\phi -4aV(\phi)\tau_t - 4aV(\phi)\tau_a \dot{a} \\
&- 4aV(\phi)\tau_\phi \dot{\phi} +4a\dot{\phi}^2 \tau_t +4a\dot{\phi}^2 \tau_a \dot{a} + 4a\dot{\phi}^3 \tau_\phi - \tau_{tt} \dot{a} - 2\tau_{ta} \dot{a}^2  \\
&- 2\tau_{t\phi }\dot{\phi}\dot{a} - 2\tau_{a\phi} \dot{a}^2 \dot{\phi} - \tau_{aa} \dot{a}^3 - \tau_{\phi\phi} \dot{\phi}^2 \dot{a} -\tau_a \dot{a}2aV(\phi)+\tau_a \dot{a}2a\dot{\phi}^2 \\
&+ 3\tau_\phi \frac{\dot{a}^2 \dot{\phi}}{a} +\tau_\phi \dot{a} \frac{dV}{d\phi}=2AV(\phi)+2a\Phi\frac{dV}{d\phi}-2A\dot{\phi}^2 \\
&-4a\dot{\phi}\Phi_t -4a\dot{\phi}\Phi_a \dot{a} - 4a\dot{\phi}^2 \Phi_\phi + 4a\dot{\phi}^2\tau_t +4a\tau_a \dot{\phi}^2 \dot{a} +4a\tau_\phi \dot{\phi}^3\\ 
\end{split}
\end{equation} 
Here we follow the same method as in [9]. On equating the coefficients of $\dot{a}^3$ we get $\tau_{aa}
=0$, thus 
\begin{equation}
\tau(t,a,\phi)=c_1(t,\phi)+c_2 (t,\phi)a 
\end{equation}
From the terms of $\dot{a}^2 \dot{\phi}$ we obtain 
\begin{equation}
-2\tau_{a\phi} +3\frac{\tau_\phi}{a}=0
\end{equation}
which using Equation (21) and after some algebra gives 
\begin{equation}
\tau =c_1 (t)+c_2 (t)a
\end{equation}

The terms of $\dot{a} \dot{\phi}^2$ give $2a\tau_a =\tau_{\phi\phi}$, thus from the previous equation we infer 
that $\tau_a =0$, so $\tau$ depends only on $t$. 
Checking the coefficients of $\dot{a}^2$ we get  
\begin{equation}
A =c_3 (t,\phi)+c_4 (t,\phi)a
\end{equation}
From the terms of $\dot{\phi}^2$ and using Equation (24) we obtain: 
\begin{equation}
\frac{\partial^2 c_3}{\partial \phi^2}+2c_3 +\frac{\partial^2 c_4}{\partial \phi^2}a+4a\Phi_\phi =0
\end{equation}
Similarly from the coefficients of $\dot{a}\dot{\phi}$ and using again Equation (24), after some algebra we 
end up with the following relation: 
\begin{equation}
a\frac{\partial c_4}{\partial \phi}+3\frac{\partial c_3}{\partial \phi} =4a^2 \Phi_a
\end{equation}
On differentiating Equation (25) with respect to $a$ we get 
\begin{equation}
a\frac{\partial^2 c_4}{\partial \phi^2}+4a\Phi_\phi +4a^2 \Phi_{a\phi} =0
\end{equation}
Similarly from equation (26) by differentiation with respect to $\phi$ we have: 
\begin{equation}
a\frac{\partial^2 c_4}{\partial \phi^2}+3\frac{\partial^2 c_3}{\partial \phi^2} =4a^2 \Phi_{a\phi}
\end{equation}
Equations (27) and (28) give: 
\begin{equation}
2a\frac{\partial^2 c_4}{\partial \phi^2}+3\frac{\partial^2 c_3}{\partial \phi^2}a +4a \Phi_\phi =0
\end{equation}
which combined again with Equation (25) implies 
\begin{equation}
a\frac{\partial^2 c_4}{\partial \phi^2} = -2\frac{\partial^2 c_3}{\partial \phi^2} +2c_3
\end{equation}
This last equation gives the functions $c_3 (t,\phi)$ and  $c_4 (t,\phi)$ in the form: 
\begin{equation}
c_3 (t,\phi)=h(t) e^\phi +k(t)e^{-\phi}
\end{equation}
\begin{equation}
c_4 (t,\phi)=f(t) +g(t)\phi
\end{equation}
where at the moment $h$, $k$, $f$ and $g$ are arbitrary functions of time. Substituting (31) and (32) back in Equation (25) 
we get for the function $\Phi(t,a,\phi)$ the following result 
\begin{equation}
\Phi =-\frac{3h(t)}{4a}e^\phi +\frac{3k(t)}{4a}e^{-\phi} +\rho(a,t)
\end{equation} 
Finally Equation (26) combined with (31), (32) and (33) gives: 
\begin{equation}
\rho(a,t) =\frac{1}{4}g(t)\ln{a}+\mu
\end{equation} 
Here we put together what we have found up to now and we get the expressions below for the functions $A(t,a,\phi)$, 
$\Phi(t,a,\phi)$ and $\tau(t,a,\phi)$: 
\begin{equation}
A =he^\phi +ke^{-\phi} +[f(t)+g\phi]a
\end{equation} 
\begin{equation}
\Phi =-\frac{3}{4a}he^\phi +\frac{3}{4a}ke^{-\phi} +\frac{1}{4}g\ln{a}+\mu
\end{equation} 
\begin{equation}
\tau =\tau(t)
\end{equation} 
Next from the coefficients of $\dot{a}$, $\dot{\phi}$ and after some algebra we end up with the relations: 
\begin{equation}
\dot{g}=\dot{h}=\dot{k}=\dot{\mu}
\end{equation} 
and 
\begin{equation}
2\dot{f}=\ddot{\tau}
\end{equation} 
The terms that do not involve derivatives of $a$ or $\phi$ give
\begin{equation}
A_{tt} +2aV(\phi)A_a -\frac{dV}{d\phi}A_\phi -4aV(\phi)\dot{\tau}=2AV(\phi)+2a\Phi\frac{dV}{d\phi}
\end{equation} 
Substituting (35), (36) and (37) into the last relation and looking at the terms that do not depend on $a$ 
after some algebra we get $h=k=0$. Similarly, the term proportional to $a\ln{a}$ gives $g=0$ and 
from the terms linear in $a$ we end up with the relation 
\begin{equation}
\ddot{f}-4V(\phi)\dot{\tau}=2\frac{dV}{d\phi}\mu 
\end{equation} 
From the last relation we infer that, if $\frac{dV}{d\phi}=0$, or equivalently if $V(\phi)$ is proportional 
to $e^{-2\phi}$ we must have $\dot{\tau}=\mu$ and $\ddot{f}=0$, from where we get 
$\tau=\mu t+c$ and $f(t)=c_1 t+c_2$. Then using Equations (38) and (39) we get $c_1 =0$. Putting all these together 
the above equations (35) through (37) become: 
\begin{equation}
A=c_1 a
\end{equation} 
\begin{equation}
\Phi =\mu
\end{equation} 
\begin{equation}
\tau=\mu t+c_2
\end{equation} 
when $V(\phi)=ce^{-2\phi}$. 
On the other hand if $V(\phi) \neq ce^{-2\phi}$ Equation (23) gives $\ddot{f}=\dot{\tau}=\mu=0$ and using 
equations (38) and (39) we get $\dot{f}=0$. So we have
\begin{equation}
A=c_1 a
\end{equation} 
\begin{equation}
\Phi=0
\end{equation} 
 \begin{equation}
\tau=c_2
\end{equation}

 
It is straight forward to prove that the vector fields determined by the Equations (42)-(44) and (45)-(47) 
both satisfy the condition just mentioned, i.e., they are symmetries of the Klein-Gordon equation. So 
we have the following theorem. 
\begin{thm} 
\label{thm} 
The Lie point symmetries of the ordinary differential Equations (12) and (13) with $V(\phi)=ce^{-2\phi}$ 
is a three dimensional Lie group with generator 

$G=(\mu t+c_2 )\frac{\partial}{\partial t} + c_1 a \frac{\partial}{\partial a} + \mu \frac{\partial}{\partial \phi}$ 

On the other hand if $V(\phi)$ arbitrary and $\neq ce^{-2\phi}$ we get a two dimensional Lie group generated by 

$c_2 \frac{\partial}{\partial t} +c_1  a \frac{\partial}{\partial a}$ 

which is a subgroup of the previous one.
\end{thm}
It is interesting to note that in the case where the self interacting potential $V(\phi)$ is exponentially decaying the 
Hartle-Hawking boundary conditions in quantum cosmology take an especially simple form [7]. We do not know 
if there is any connection between this result and the above theorem , where again the exponentially decaying  potential  
plays a special role. 
The quantity $E\equiv \dot{a}^2 -2a^2 V(\phi) -a^2 \dot{\phi}^2$, 
which is $-k$  from the Einstein's constraint equation, 
is preserved by the solutions of the above Equations (12) and (13).  
We can easily find how the above vector fields act on $E$: 
\begin{equation} 
pr^{(1)} G(E)=2(c_1 -\mu)E
\end{equation}
This shows that in the case of a flat universe $G$ is a symmetry group for all Einstei's Equations (2), (3) 
and (5). The generators of the symmetry group $G$ are given by: 
\begin{equation} 
\vec{X}=t\frac{\partial}{\partial t} +\frac{\partial}{\partial \phi}
\end{equation}
\begin{equation} 
\vec{Y}=\frac{\partial}{\partial t}
\end{equation}
\begin{equation} 
\vec{Z}=a\frac{\partial}{\partial a} 
\end{equation}
The vector field $\vec{X}$ corresponds to $c_1 =c_2 =0$, $\mu \neq 0$, the vector field $\vec{Y}$ 
corresponds to $c_1 =\mu =0$, $c_2 \neq 0$ and 
$\vec{Z}$  corresponds to $c_2 =\mu =0$, $c_1 \neq 0$. Their multiplication table is 
\begin{equation}
[\vec{X} ,\vec{Y}]=-\vec{Y}
\end{equation}
\begin{equation}
[\vec{X} ,\vec{Z}]=0
\end{equation}
\begin{equation}
[\vec{Y} ,\vec{Z}]=0
\end{equation}
So here we have a solvable but not nilpotent Lie group. For closed universes and for open hyperbolic only, $\vec{Y}$ 
is a symmetry for all Einstein's equations. In the special case of $V(\phi)=ce^{-2\phi}$ 
the subgroup generated by $\vec{Y}$ and $\vec{W}=\vec{X}+\vec{Y}$ leaves invariant all Einstein's equations for all 
the cases where $k=-1,0,+1$. 

In a recent study of the Maxwell-Bloch system [3], we found again only solvable Lie groups of symmetry. 
We do not know if there is any general condition which both systems satisfy and which forces them 
to accept only solvable groups of symmetry. This has to be investigated. 

\section 
{Reduction} 

It is well known that if a one parameter Lie group of transformations is admitted by an ordinary differential 
equation then its order can be reduced by one. For  first 
order ordinary differential equations this corresponds to a reduction to quadrature. This reduction of order 
can always be accomplished by using canonical 
ccordinates accosiated with the group. For higher order ordinary differential equations the reduction 
in order can be accomplished by using differential invariants. Here we will reduce the second order system 
consisting of the Equations (12) and (13) to a first order system using the change of variables: 
\begin{equation}
\phi \equiv x
\end{equation}  
\begin{equation}
\dot{\phi} \equiv y
\end{equation}
\begin{equation}
\frac{\dot{a}}{a} \equiv w
\end{equation}
which are invariants of the symmetry group generated by $\vec{Y}$ and $\vec{Z}$. It is trivial to show that under 
the above transformation the second order system (12)-(13) reduces 
to the following first order system of ordinary differential equations 
\begin{equation}
y\frac{dy}{dx}=-3wy-\frac{dV}{dx}
\end{equation}
\begin{equation}
y\frac{dw}{dx}+w^2 =2V(x)-2y^2
\end{equation}
and two quadratures. If we find $y(x)$ and $w(x)$ then we can find $\phi(t)$ and $a(t)$ by quadratures: 
\begin{equation}
\int \frac{d\phi}{y(\phi)}=t
\end{equation}
Inverting the relation above we can get $\phi$ as a function of $t$. Then we can find $a$ as a function of $t$ using the quadrature 
\begin{equation}
a=\exp \int w(\phi(t))dt
\end{equation}
Unfortunately the reduced system consisting of the Equations (59) and (60) is not autonomous even though the original one is. 
Using $x$, $y$ and $w$ the quantity $E$ can be written in the form 
\begin{equation}
E=a^2 \Big[w^2 -2V(x) -y^2 \Big]
\end{equation}
and it is trivial to prove that the total derivative of $E$ with respect to $x$ is zero. So $E$ is still conserved 
even though is not expressed solely in terms of the new variables $x$, $y$ and $w$. 
Both Equations (12) and (13) are locally solvable and of maximal rank since their rank is the rank of the matrix 
\[ \left ( \begin{array}{cccccc} 
-2V +2\dot{\phi}^2 & -2a\frac{dV}{d\phi} & 0 & 4a\dot{\phi} &1 &0 \\ 
-3\frac{\dot{a}\dot{\phi}}{a^2 } & \frac{d^2 V}{d\phi^2 } & 3\frac{\dot{a}}{a} & 0 & 0 & 1
\end{array} \right ) \]
As a result, conditions (16) and (17) are necessary and sufficient for $G$ to be a Lie point symmetry.  

\section 
{Variational symmetries} 

Even though Equations (2), (3) and (5) can be reduced from a variational principle, as we have seen in the 
introduction, the study of 
variational symmetries is problematic and the application of Noether's  theory to this specific 
variational process, to our knowledge, is an open problem. 

We will look here more closely at the peculiarities of our variational problem. As we have seen in the introduction, 
the original variational equations are Equations (2), (3) and (4). We have replaced Equation (4) by Equation (5) since combining 
algebraically Equations (2) and (4) we get (5). 
Thus we can use Equations (2), (3) and (5), which are equivalent with  the original equations. 
These are three equations for the three uknown functions 
$N(t)$, $a(t)$ and $\phi(t)$. But actually they are not independent, because using (2) and (3) we can prove 
(5); using (2) and (5) we can prove (3) and even using (3) and (5) we can prove (2) 
(but we can not infer the value of $k$). On the other hand if in the above equations we rescale the time, 
by using as our new time the integral of the lapse function, i.e., the proper time of general relativity, we will 
have three (dependent) equations for only two unknown functions, $a(t)$ and $\phi(t)$, where $t$ now stands for 
the proper time. The resulting equations are not in any obvious way the Euler-Lagrange equations 
of a variational problem and from here it springs the difficulty of applying Noether's theory. 
It looks that $N$ is a quantity we need to formulate the problem but eventually disappears from the dynamics.  
it is like an ignorable quantity. This is related with the problem of distinguishing the gauge and 
the dynamics in general relativity a main source of difficulties for both classical and quantum general relativity. 
See for example the articles of Teitelboim [13], Kushar and  [8], Hanson, Regge 
and Teitelboim [6].     

Here we will use from the beginning the proper time in the Einstein-Hilbert action and we will study the 
resulting variational problem. The action takes the form 
\begin{equation}
S=\frac {1}{2}\int dt \Big[ -a \Big(\frac {da}{dt}\Big)^2  +ka +a^3 \Big( \frac {d \phi}{dt}\Big)^2 - 2a^3 V \Big]
\end{equation}
where $t$ from now on will represent the proper time. Varying with respect to $\phi$ and $a$ we get the 
Equations (3) and (4) of the introduction. We do  not have here the 
Hamiltonian constrained equation and as a result we do not have also Equation (5). 
In short the resulting equations are not equivalent with the system we had before. 
Following the same method as in Section 2 we get the following result: 
\begin{thm} 
The Lie point symmetries of the Equations (3) and (4) with $k=0$ is a two dimensional solvable Lie group with generators 
$G=c_2  \frac{\partial}{\partial t} + c_1 a \frac{\partial}{\partial a} $ 
For $k \neq 0$ the only symmetry is time translation. The above hold independently of the form of the potential $V(\phi)$.
\end{thm}
Equations (3) and (4) are the Euler-Lagrange equations for the Lagrangian 
$L=-\frac{1}{2} a \dot{a}^2 + \frac{1}{2} ka + \frac{1}{2} a^3 \dot{\phi}^2 -a^3 V$ 
We can easily prove that the time translation is a variational symmetry, since it satisfies the condition 
$pr^{(1)} \vec{X}(L)+L\vec{\nabla} \cdot \vec{\xi}=0$  
which is necessary and sufficient condition for a vector field $\vec{X}$ to be a variational symmetry. 
On the other hand scaling in $a$ is not a variational symmetry. From Noether's theory 
we can find the conserved  quantity associated with time translation: 
If $\vec{X}=\sum \limits_{i=1}^p {\lambda}^i \frac{\partial}{\partial t^i} + 
\sum \limits_{\alpha =1}^q \psi_\alpha (t,u) \frac{\partial}{\partial u^\alpha }$ 
is a variational symmetry, where $t^i$ are the independent and $u^\alpha$ are the dependent variables. 
Then the characteristics of $\vec{X}$ are given by 
$Q_\alpha =\psi_\alpha -  
\sum \limits_{i=1}^p \lambda^i \frac{\partial u^\alpha }{\partial t^i }$
and they generate a conservation law in characteristic form by the equation 
$\vec{\nabla} \cdot \vec{P} = \sum \limits_{\alpha =1}^q Q_\alpha \cdot E_\alpha (L)$ 
For the variational symmetry $\frac{\partial}{\partial t}$ we can easily find its characteristics 
$Q_a = -\dot {a}$ and $Q_\phi =-\dot{\phi}$. The conservation law takes the form 
$\frac{dP}{dt} = Q_a \cdot E_a (L)+ Q_\phi \cdot E_\phi (L)$. 
where $L$ is the above Lagrangian and $E_a$, $E_\phi$ are the Euler-Lagrange operators 
corresponding to $a$ and $\phi$. Substituting from Equation (63) we get:
\begin{equation}
Q_a \cdot E_a (L)+ Q_\phi \cdot E_\phi (L) = \frac {1}{2}\frac {d}{dt} \Big[ a(\dot{a}^2  +ka -2a^3 V-a^3 \dot{\phi}^2 \Big]
\end{equation}
So $P=a \cdot E$ is a conserved quantity for Equations (3) and (4), where $E$ is the conserved quantity 
we have for Equations (12) and (13). We could find the above conserved quantity 
$P$ by considering $t$ as a function of either $a$ or $\phi$. If $a$ is our independent variable the action takes the form 
\begin{equation}
S=\frac{1}{2} \int dat_a \Big[ -\frac{a}{t_{a}^2}  +ka+a^3 \frac{\phi_{a}^2}{{t_a}^2} - 2a^3 V \Big], 
\end{equation}
the  Lagrangian does not depend on time, as we would expect, and the conjugate momentum 
$\frac{\partial L}{\partial t_a}$ is conserved. It is trivial to check that this conserved quantity agrees with $P$. 
Actually this is the Hamiltonian related to the Lagrangian above, something we should expect since 
it is related to the time translation symmetry. 

For $k=0$ we get the same symmetry group as in sections two and three, so we can use as  new 
variables the invariants $x$, $y$ and $z$ of section three to reduce the order of the Equations (3) and (4). 
Indeed, in terms of $x$, $y$ and $z$  they take the form 
\begin{equation}
y\frac{dy}{dx} +3wx+\frac{dV}{dx}=0 \nonumber
\end{equation}
\begin{equation} 
2y\frac{dw}{dx} +3w^2 +3y^2 -6V(x) =0 \nonumber
\end{equation} 

\section 
{Variational symmetries for the full system} 

In the original action the Lagrangian is written in terms of $a$, $\phi$, $N$ and $t$, which is the affine time, in the form 
\begin{equation}
L=-\frac{1}{2} \frac{a}{N} \dot{a}^2 +\frac{1}{2} kNa+\frac{1}{2} \frac{a^3 \dot {\phi}^2}{N} -Na^3 V(\phi)
\end{equation}
The equations of motion for $a$,  $\phi$ and $N$ take now the form 
\begin{equation}
\dot{a}^2 +N^2 k -a^2 \dot{\phi}^2 -2N^2 a^2 V=0
\end{equation}
\begin{equation}
aN\ddot{\phi} +3N\dot{a} \dot{\phi} -a\dot{N} \dot{\phi} +N^3a\frac{dV}{d\phi}=0
\end{equation}
and 
\begin{equation}
N\ddot{a} +2Na\dot{\phi}^2 -\dot{N}\dot{a}-2N^3 aV=0)
\end{equation}
We can easily verify that the vectors $\vec{X}$, $\vec{Y}$, $\vec{Z}$ of section three are still 
Lie point symmetries for  Equations (66), (67) and (68), i.e., 
they satisfy the following conditions  
\begin{equation}
pr{^(2)} \vec{X} (Eq1)=0 \nonumber
\end{equation}
for $k=0$ and $V(\phi)=c\exp(-2\phi)$ and 
\begin{equation}
pr^{(2)} \vec{X} (Eq2)=pr^{(2)} \vec{X} (Eq3)=0 \nonumber
\end{equation}
for arbitrary $V(\phi)$ and $k$. Also, 
\begin{equation}
pr^{(2)} \vec{Y} (Eq1, Eq2, Eq3)=0 \nonumber
\end{equation}
for any $V(\phi)$ and $k$. And finally, 
\begin{equation}
pr^{(2)} \vec{Z} (Eq1)=0 \nonumber
\end{equation}
for any $V(\phi)$ but $k=0$ and 
\begin{equation}
pr^{(2)} \vec{Z} (Eq2, Eq3)=0 \nonumber
\end{equation}
for any $V(\phi)$ and $k$, where now 
\begin{equation}
pr^{(2)} \vec{X} =t\partial _t +\partial _\phi -\dot{a} \partial _{\dot{a}} -\dot{\phi}\partial _{\dot{\phi}} 
-\dot{N} \partial _{\dot{N}} -2\ddot{a} \partial _{\ddot{a}} -2\ddot{\phi} \partial _{\ddot{\phi}} 
- 2\ddot{N} \partial _{\ddot{N}}\nonumber  
\end{equation}
\begin{equation} 
pr^{(2)} \vec{Y} =\partial_t  \nonumber
\end{equation}
\begin{equation}
pr^{(2)} \vec{Z} =a\partial _a +\dot{a} \partial _{\dot{a}}  +\ddot{a} \partial _{\ddot{a}}  \nonumber
\end{equation}
Here we can also check which of the above vector fields are variational symmetries. It turns out that $\vec{X}$ 
and $\vec{Z}$ are not variational symmetries, but  $\vec{Y}$ is a variational symmetry 
satisfying the condition 
\begin{equation}
pr^{(1)} \vec{Y}(L)+L\vec{\nabla} \cdot \vec{\xi}=0  \nonumber
\end{equation} 
The characteristics of $\vec{Y}$ are $Q_a=-\dot{a}$, $Q_\phi=-\dot{\phi}$, $Q_N=-\dot{N}$ and 
applying again Noether's theorem we get the corresponding 
conservation law in the form 
\begin{equation}
\frac{dK}{dt}=Q_aE_a(L)+Q_\phi E_\phi(L)+Q_NE_N(L)  \nonumber
\end{equation} 
which gives
\begin{equation}
\frac{dK}{dt}= \frac{d}{dt} \Big[\frac{a}{2N} \Big(\dot{a}^2 + N^2 k -2N^2 a^2 V -a^2 \dot{\phi}^2 \Big) \Big] \nonumber 
\end{equation}
This agrees with the conserved quantity $P$ of section four if we set $N=1$. Of course $K$ does 
not give a new conservation law since using Equation (66) we get $K=0$.

\end{document}